\documentclass[12pt]{article}

\input {tcilatex}

\begin{document}

\author{Zakaria Giunashvili \\
A. Razmadze Mathematics Institute\\
Georgian Academy of Sciences\\
Department of Theoretical Physics\\
e-mail: {\bf zaqro@gtu.edu.ge}}
\title{Differential Complex of Poisson Manifold and Distributions}
\date{11 March, 2001}
\maketitle

\begin{abstract}
We study variuos homological structures associated with Poisson algebra, the
canonical differential complex for singular Poisson structure and the
analogue of the star operator for such manifolds.

Give the interpretation of the classical Koszul differential of exterior
forms, as the supercommutator with some second order element.

Describe the space of invariant distributions on manifold with singular
Poisson structure.
\end{abstract}

\footnotetext{Supported by INTAS international grant}

\section{Lie superalgebra structure on the space of multiderivations of a
commutative algebra. Poisson cohomologies}

Let $A$ be a real or complex vector space. For each integer $k$ let $L^k(A)$
be the space of multilinear antisymmetric maps from $A^k$ into $A$. Let $%
L^0(A)$ be $A$ and $L(A)$ be $\oplus _{k=0}^\infty L^k(A)$.

There is a natural Lie algebra structure on the space $L^1(A)$ defined by
the commutator of the composition. The composition on this space can be
extended to the operator on $L(A)$ called the compositional product (see 
\cite{nijenhuis}):%
$$
\begin{array}{l}
for\,\,\,\alpha \in L^m(A)\,\,\,and\,\,\,\beta \in
L^n(A)\,\,\,define\,\,\,\alpha \circ \beta \in L^{m+n-1}(A)\,\,\,as \\  
\\ 
(\alpha \circ \beta )(u_1,\ldots ,u_{m+n-1})= \\  
\\ 
=\sum\Sb s(1)<\cdots <s(n) \\ s(n+1)<\cdots <s(m+n-1)\endSb sgn(s)\alpha
(\beta (u_{s(1)},\ldots u_{s(n)}),u_{s(n+1)},\ldots u_{s(m+n-1)}) 
\end{array}
$$
As a result, the commutator on $L^1(A)$ can be extended to the
supercommutator on the space $L(A)$ as 
\begin{equation}
\label{Formula0}[\alpha ,\,\,\,\beta ]=(-1)^{(m+1)n}\alpha \circ \beta
+(-1)^m\beta \circ \alpha 
\end{equation}
This bracket satisfies the conditions needed for the space $L(A)$ be a
superalgebra:

for $\alpha \in L^m(A)$, $\beta \in L^n(A)$ and $\gamma \in L^k(A)$

\begin{description}
\item[(a)]  $[\alpha ,\,\,\,\beta ]=(-1)^{mn}[\beta ,\,\,\,\alpha ];$

\item[(b)]  $(-1)^{mk}[[\alpha ,\,\,\,\beta ],\,\,\,\gamma
]+(-1)^{mn}[[\beta ,\,\,\,\gamma ],\,\,\,\alpha \bar ]+(-1)^{nk}[[\gamma
,\,\,\,\alpha ],\,\,\,\beta ]=0.$
\end{description}

An element $\mu \in L^2(A)$ satisfying the condition $[\mu ,\,\,\,\mu ]=0$,
defines a Lie algebra structure on $A$:%
$$
for\,\,\,a,b\in A\,\,\,let\,\,\,[a,\,\,\,b]_\mu =\mu (a,\,\,\,b) 
$$
We call such element {\em involutive}.

An involutive element $\mu \in L^2(A)$ defines a linear operator\\$\partial
_\mu :L(A)\longrightarrow L(A)$, of degree $+1$: $\partial _\mu (\alpha
)=[\mu ,\,\,\,\alpha ]$. From the condition (b) and the involutiveness of
the element $\mu $ follows that the operator $\partial _\mu $ is a
coboundary operator, i.e. $\partial _\mu \circ \partial _\mu =0$.

Let $A$ be a commutative algebra over the field of the real or complex
numbers. In this case, the space $L(A)$\thinspace has a structure of an
exterior algebra under the multiplication operator defined by the classical
formula 
\begin{equation}
\label{Formula1}
\begin{array}{l}
for\,\,\,\alpha \in L^m(A),\,\,\,and\,\,\,\beta \in L^n(A),\,\,\,let \\  
\\ 
(\alpha \wedge \beta )(u_1,\ldots ,u_{m+n})= \\ 
=\frac 1{m!n!}\sum_ssgn(s)\alpha (u_{s(1)},\ldots ,u_{s(m)})\beta
(u_{s(m+1)},\ldots ,u_{s(m+n)}) 
\end{array}
\end{equation}
We call an element $\alpha \in L(A)$ {\em multiderivation} if for any set of
elements\\$a,\,\,\,a_1,\ldots ,a_m\in A:$%
$$
\alpha (aa_1,a_2,\ldots ,a_k)=a\alpha (a_1,\ldots ,a_k)+a_1\alpha
(a,a_2,\ldots ,a_k) 
$$

The subspace of all multiderivations in $L^m(A)$ we denote by $Der^m(A)$.
Also, we put : $Der^0(A)=A$ and $Der(A)=\oplus _{k=0}^\infty Der^k(A)$.

The subspace $Der(A)$ in the space $L(A)$ is closed as under the operator of
exterior multiplication defined by the formula \ref{Formula1}, so under the
bracket defined by the formula \ref{Formula0}. Moreover, these two
structures are interconnected by the following property: 
\begin{equation}
\label{Formula2}
\begin{array}{l}
for\,\,\,\alpha \in Der^m(A),\,\,\,\beta \in Der^n(A)\,\,\,and\,\,\gamma \in
Der(A) \\  
\\ 
we\,\,\,have \\  
\\ 
\lbrack \alpha ,\,\,\,\beta \gamma ]=[\alpha ,\,\,\,\beta ]\wedge \gamma
+(-1)^{(m+1)n}\beta \wedge [\alpha ,\,\,\,\gamma ] 
\end{array}
\end{equation}

For any integer $k$, we have the subspace $\wedge ^kDer^1(A)$ in $Der^k(A)$,
which is the set of the elements of the type $v_1\wedge \ldots \wedge v_k$,
where each $v_i,\,\,\,i=1,\ldots ,k$ is an element of the space $Der^1(A)$.
The subalgebra $\wedge ^kDer^1(A)$ in the algebra $Der(A)$ is also closed
under the bracket $[\,\,\,,\,\,\,]$ and the exterior multiplication. For the
restriction of the bracket on the algebra $\wedge ^kDer^1(A)$, the following
explicit formula can be used: 
\begin{equation}
\label{Formula3}
\begin{array}{l}
\lbrack u_1\wedge \ldots \wedge u_m,\,\,\,v_1\wedge \ldots \wedge v_n]= \\ 
\sum_{i,j}(-1)^{m+i+j-1}[u_i\,\,\,,\,\,\,v_j]\wedge u_1\wedge \ldots \wedge 
\widehat{u_i}\wedge \ldots \\ \ldots \wedge u_m\wedge \,v_1\wedge \ldots
\wedge \widehat{v_j}\wedge \ldots \wedge v_n 
\end{array}
\end{equation}
where $u_1,\ldots ,u_m$ and $v_1,\ldots ,v_n$ are the elements of the space $%
Der^1(A)$.

An involutive element $P\in Der^2(A)$ defines a bracket on $A$ which, at the
same time, is a biderivation. Such a structure on a commutative algebra is
called the {\em Poisson structure}. As the subspace $Der(A)$ is closed under
the supercommutator, it will be invariant under the action of the operator $%
\partial _P$, and therefore we have the subcomplex $(Der(A),\,\,\,\partial
_P)$ of the complex $(L(A),\,\,\,\partial _P)$. From the formula \ref
{Formula2} for the bracket $[\,\,\,,\,\,\,]$, on $Der(A)$, follows that the
operator $\partial _P:Der(A)\longrightarrow Der(A)$ is an antidifferential,
that is%
$$
\begin{array}{l}
for\,\,\,each\,\,\,u\in Der^m(A)\,\,\,and\,\,\,v\in
Der^n(A)\,\,\,we\,\,\,have \\  
\\ 
\partial _P(u\wedge v)=\partial _P(u)\wedge v+(-1)^mu\wedge \partial _P(v) 
\end{array}
$$
Therefore, on the cohomologies of the complex $(Der(A),\,\,\,\partial _P)$
can be induced the structure of exterior algebra from $Der(A)$. The
cohomology algebra of the complex $(Der(A),\,\,\,\partial _P)$ is called the 
{\em cohomology algebra of the Poisson algebra} $(A,\,\,\,P)$.

For the commutative algebra $A$, consider the space of the first order
differential operators from $A$ to itself, denoted by $Diff^1(A)$. By
definition, $Diff^1(A)$ is the subspace of the space $Hom(A,\,\,\,A)$
consisting of the mappings $\varphi :A\longrightarrow A$, such that, for any 
$a\in A=Hom_A(A,\,\,\,A)\subset Hom(A,\,\,\,A)$, we have that $[\varphi
,\,\,\,a]\in Hom_A(A,\,\,\,A)=A$.

As it is well-known $Diff^1(A)=Der^1(A)\oplus A$:

$\varphi (uv)-u\varphi (v)=(\varphi -\varphi (1))(uv)-u(\varphi -\varphi
(1))(v)$

So, we have that $(\varphi -\varphi (1))(uv)-u(\varphi -\varphi
(1))(v)=c(u)v $. Consequently, putting $v=1$ we get $c=\varphi -\varphi (1)$%
, and therefore, the mapping%
$$
X=\varphi -\varphi (1):A\longrightarrow A 
$$
is an element of the space $Der^1(A)$, and any element $\varphi \in
Diff^1(A) $, can be decomposed as $\varphi =(\varphi -\varphi (1))+\varphi
(1)$, where $\varphi -\varphi (1)\in Der^1(A)$ and $\varphi (1)\in A$.

The anticommutative algebra structure described above, can be considered as
well in the case of $L(B)$, where $B=Diff^1(A)$. The space $B$ is equipped
with the natural structure of Lie algebra, defined by the commutator $%
[u,\,\,\,v]=u\circ v-v\circ u$ which means that there exists an involutive
element $\mu \in L^2(B)$. So, we can consider the complex $%
(L(Diff^1(A)),\,\,\,\partial _\mu )$.

For each positive integer $n$ let $\Omega ^n(A)$ be the subspace of the space%
\\$L^n(Diff^1(A))$, consisting of such mappings%
$$
\omega :Diff^1(A)\times \ldots \times Diff^1(A)\longrightarrow Diff^1(A) 
$$
that

\begin{itemize}
\item  $\omega $ takes values in $A=Hom_A(A,\,\,\,A)\subset Diff^1(A)$;

\item  $\omega (u_1,\ldots u_n)=0$ if at least one of the elements $%
u_1,\ldots u_n\in Diff^1(A)=Der^1(A)\oplus A$ is in $A$;

\item  $\omega $ is an $A-multilinear$, i.e. $\omega (au_1,\ldots
,u_n)=a\omega (u_1,\ldots ,u_n)$, for any $a\in A$ and $u_1,\ldots ,u_n\in
Der(A)$.
\end{itemize}

Using the classical terminology, it can be said that the elements of the
subspace $\Omega ^n(A)$ are differential forms of the order $n$ on the
commutative algebra $A$.

\begin{theorem}
The subspace $\Omega (A)=\oplus _{n=0}^\infty \Omega ^m(A)$ in the space $%
L(Diff^1(A))$ is invariant under the action of the operator $\partial _\mu
=[\mu ,\,\,\,\cdot ]$, and the restriction of the operator $-\partial _P$ on 
$\Omega (A)$ coincides with the classical differential on space of
differential forms.
\end{theorem}

{\bf Proof. }By the definition of the supercommutator, for $\omega \in
\Omega ^n(A)$ we have the following%
$$
\begin{array}{l}
\lbrack \mu ,\,\,\,\omega ](u_1,\ldots ,u_{n+1})=(-1)^n\sum (-1)^{n+1-i}\mu
(\omega (u_1,\ldots , 
\widehat{u}_i,\ldots ,u_{n+1}),\,\,\,u_i)+ \\ +\sum (-1)^{i+j-1}\omega (\mu
(u_i,u_j),\,\,u_1,\ldots , 
\widehat{u}_i,\ldots ,\widehat{u}_j,\ldots ,u_{n+1})= \\ =\sum
(-1)^{i-1}[\omega (u_1,\ldots , 
\widehat{u}_i,\ldots ,u_{n+1}),\,\,\,u_i]+ \\ +\sum (-1)^{i+j-1}\omega
([u_i,\,\,\,u_j],\,u_1,\ldots , 
\widehat{u}_i,\ldots ,\widehat{u}_j,\ldots ,u_{n+1})= \\ =\sum
(-1)^iu_i\omega (u_1,\ldots , 
\widehat{u}_i,\ldots ,u_{n+1})+ \\ +\sum (-1)^{i+j-1}\omega
([u_i,\,\,\,u_j],\,u_1,\ldots , 
\widehat{u}_i,\ldots ,\widehat{u}_j,\ldots ,u_{n+1})= \\ =-(\sum
(-1)^{i-1}u_i\omega (u_1,\ldots , 
\widehat{u}_i,\ldots ,u_{n+1})+)+ \\ +\sum (-1)^{i+j}\omega
([u_i,\,\,\,u_j],\,u_1,\ldots , 
\widehat{u}_i,\ldots ,\widehat{u}_j,\ldots ,u_{n+1})= \\ =-(d\omega
)((u_1,\ldots ,u_{n+1})) 
\end{array}
$$

\bigskip\ 

To summarize, we can state that the subspace $\Omega (A)$ in the space\\$%
L(Diff^1(A))$ is not closed under the operation of supercommutator, but it
is invariant under the action of the operator $[\mu ,\,\,\,\cdot ]$, where $%
\mu \in L^2(Diff^1(A))$ is the element defined by the commutator in the Lie
algebra $Diff^1(A)$. It can be defined the operation of the exterior
multiplication in the space $\Omega (A)$ by the formula \ref{Formula1},
after which the operator $d$ becomes the antiderivation of degree $+1$ of
the algebra $\Omega (A)$.

Any element $p\in Der^2(A)$ defines the mapping $\widetilde{p}%
:A\longrightarrow Der^1(A)$ as follows%
$$
\widetilde{p}(a)(b)=p(a,\,\,b) 
$$
which can be extended to the mapping $\widetilde{p}:\Omega
(A)\longrightarrow Der(A)$ by the following formula%
$$
\widetilde{p}(\alpha )(a_1,\ldots ,a_n)=(-1)^n\alpha (\widetilde{p}%
(a_1),\ldots ,\widetilde{p}(a_n)) 
$$
where $\alpha \in \Omega (A)$ and $a_1,\ldots ,a_n\in A$.

As it was mentioned, the involutieness of the element $p$ (i.e. $%
[p,\,\,\,p]=0$) is equivalent to the bracket $\{a,\,\,\,b\}=p(a,\,\,\,b)$ be
a Lie algebra structure on $A$:%
$$
\begin{array}{l}
\lbrack p,\,\,\,p](a,b,c)=2(p(p(a,b),c)+p(p(b,c),a)+p(p(c,a),b))= \\ 
=2(\{\{a,b\},c\}+\{\{b,c\},a\}+\{\{c,a\},b\}) 
\end{array}
$$

\begin{lemma}
If $p$ is involutive, the mapping $\widetilde{p}:A\longrightarrow Der^1(A)$
is a Lie algebra homomorphism.
\end{lemma}

{\bf Proof. }$\widetilde{p}(\{a,b\})(c)=\{\{a,b\},c\}$; as it follows from
the Jacoby identity for the bracket $\{\}$, we have%
$$
\{\{a,b\},c\}=\{a,\{b,c\}\}-\{b,\{a,c\}\}=(\widetilde{p}(a)\widetilde{p}(b)-%
\widetilde{p}(b)\widetilde{p}(a))(c) 
$$

\bigskip\ 

\begin{theorem}
The mapping $\widetilde{p}:\Omega (A)\longrightarrow Der(A)$ is a
homomorphism of the complexes $(\Omega (A),\,\,\,d)$ and $%
(Der(A),\,\,\,\partial _p=[p,\,\,\,\cdot ])$.
\end{theorem}

{\bf Proof. }$\widetilde{p}(d\omega )(a_1,\ldots
,a_{n+1})=(-1)^{n+1}(d\omega )(\widetilde{p}(a_1),\ldots ,\widetilde{p}%
(a_{n+1}))=$

$=(-1)^{n+1}(\sum_i(-1)^{i-1}\widetilde{p}(a_i)\omega (\widetilde{p}%
(a_1),\ldots ,\widehat{\widetilde{p}(a_1)},\ldots ,\widetilde{p}(a_{n+1}))+$

$+\sum_{i<j}(-1)^{i+j}\omega ([\widetilde{p}(a_i),\,\,\widetilde{p}%
(a_j)],\ldots ,\widehat{\widetilde{p}(a_i)},\ldots ,\widehat{\widetilde{p}%
(a_j)},\ldots ,\widetilde{p}(a_{n+1})))=$

$=(-1)^{n+1}(\,\,\sum_i(-1)^{i-1}p(a_i,\,\,\omega (\widetilde{p}(a_1),\ldots
,\widehat{\widetilde{p}(a_1)},\ldots ,\widetilde{p}(a_{n+1}))+$

$+\sum_{i<j}(-1)^{i+j}\omega ([\widetilde{p}(a_i),\,\,\widetilde{p}%
(a_j)],\ldots ,\widehat{\widetilde{p}(a_i)},\ldots ,\widehat{\widetilde{p}%
(a_j)},\ldots ,\widetilde{p}(a_{n+1}))\,\,)$.

On the other hand we have:

$[p,\,\,\,\widetilde{p}(\omega )](a_1,\ldots ,a_{n+1})=\sum_i(-1)^{i-1}p((%
\widetilde{p}(\omega ))(a_1,\ldots ,\widehat{a_i},\ldots ,a_{n+1}),\,\,a_i)+$

$+\sum_{i<j}(-1)^{i+j-1}\widetilde{p}(\omega )(p(a_i,\,\,a_j),\ldots ,%
\widehat{a_i},\ldots ,\widehat{a_j},\ldots ,\widetilde{p}(a_{n+1}))=$

$=(-1)^{n+1}(\,\sum_i(-1)^{i-1}p(a_i,\,\,\omega (\widetilde{p}(a_1),\ldots ,%
\widehat{\widetilde{p}(a_i)},\ldots ,\widetilde{p}(a_{n+1}))\,)+$

$+\,\sum_{i<j}(-1)^{i+j-1}\omega ([\widetilde{p}(a_i),\,\,\widetilde{p}%
(a_j)],\ldots ,\widehat{\widetilde{p}(a_i)},\ldots ,\widehat{\widetilde{p}%
(a_j)},\ldots ,\widetilde{p}(a_{n+1}))\,\,)$

\bigskip\ 

\section{Schouten bracket as the deviation of the coboundary operator from
the Leibniz\protect\\rule}

The main result of the previous section is the fact that a supercommutator
on an exterior algebra gives rise of some coboundary operator on this
algebra, and even the classical differential on the exterior algebra of
differential forms can be represented as a supercommutator with some second
order element of some superalgebra containing the algebra of differential
forms. In this section, we consider some reverse situation: a coboundary
operator on some exterior algebra induces a superalgebra structure on this
algebra.

Let $E$ be a real or complex $Z$-graded exterior algebra with a
multiplication operation denoted by $\wedge $. Let $\partial
:E\longrightarrow E$ be a boundary operator ($\partial \circ \partial =0$
and $\partial (E_i)\subset E_{i-1}$ for $i=0,\cdots ,\infty $).

The operator $\partial $ is said to be an antidifferential if for any $u\in
E_m$ and $v\in E$, it satisfies the following condition%
$$
\partial (u\wedge v)=\partial (u)\wedge v+(-1)^mu\wedge \partial (v) 
$$
For any boundary operator on the exterior algebra $E$ we can define the
bilinear mapping $[\,\,\,,\,\,\,]:E\times E\longrightarrow E$ as follows 
\begin{equation}
\label{Formula4}
\begin{array}{l}
for\,\,\,u\in E_m\,\,\,and\,\,\,v\in E\,\,\,let \\  
\\ 
\lbrack u,\,\,\,v]=\partial (u)\wedge v+(-1)^mu\wedge \partial (v)-\partial
(u\wedge v) 
\end{array}
\end{equation}
If the operator $\partial $ is antiderivation, the mapping defined by this
formula is trivial.

In any case, we can ask the question is the bracket $[\,\,\,,\,\,\,]$ a Lie
superalgebra structure on $E$ or not? To be so, the following conditions
must be hold:

for any $u\in E_m$, $v\in E_n$ and $w\in E_k$

\begin{description}
\item[(s1)]  $[u,\,\,\,v]=(-1)^{mn}[u,\,\,\,v]$

\item[(s2)]  $[u,\,\,\,v\wedge w]=[u,\,\,\,v]\wedge
w+(-1)^{(m+1)n}\,\,\,v\wedge [u,\,\,\,w]$

\item[(s3)]  $(-1)^{mk}[[u,\,\,\,v],\,\,\,w]+(-1)^{mn}[[v,\,\,\,w],\,\,%
\,u]+(-1)^{nk}[[w,\,\,\,u],\,\,\,v]=0$
\end{description}

The first of these three conditions is obviously always true. The third one
is also always true for the first order elements and is equivalent to $%
(\partial \circ \partial )(x\wedge y\wedge z)=0$, for $x,\,\,y,\,\,z\in E_1$%
; and implies that the bracket $[x,\,\,\,y]=-\partial (x,\,\,y)$ defines a
Lie algebra structure on $E_1$.

The condition (s2) is equivalent to the following equality for the operator $%
\partial $: 
\begin{equation}
\label{FormulaX4}
\begin{array}{l}
\partial (\alpha \wedge \beta \wedge \gamma )= \\ 
=\partial (\alpha \wedge \beta )\wedge \gamma +(-1)^m\alpha \wedge \partial
(\beta \wedge \gamma )+(-1)^{(m+1)n}\beta \wedge \partial (\alpha \wedge
\gamma )- \\ 
-(\partial \alpha \wedge \beta \wedge \gamma +(-1)^m\alpha \wedge \partial
\beta \wedge \gamma +(-1)^{m+n}\alpha \wedge \beta \wedge \partial \gamma ) 
\end{array}
\end{equation}

It is easy to check by induction that the condition (s2) implies that the
operator $\partial $, on the elements of the type $u_1\wedge \ldots \wedge
u_n\in E_n$ where $u_1,\ldots ,u_n\in E_1$ has the form%
$$
\partial (u_1\wedge \ldots \wedge
u_n)=\sum_{i<j}(-1)^{i+j}[u_i,\,\,\,u_j]\wedge u_1\wedge \ldots \wedge 
\widehat{u_i}\wedge \ldots \wedge \widehat{u_j}\wedge \ldots \wedge u_n 
$$
and in this case, all of the above three conditions are true on the
subalgebra $\wedge E_1=\oplus _{k=0}^\infty (\wedge ^kE_1)$.

Let a Lie algebra $L$ be a module over some commutative real or complex $A$,
which, itself is a module over the Lie algebra $L$. That is: there is a Lie
algebra homomorphism from $L$ into the Lie algebra of all derivations of the
algebra $A$. Assume that these two structures: the $A$-module structure on $%
L $ and the $L$-module structure on $A$, are interconnected by the following
condition:

for any $x,\,\,\,y\in L$ and $a\in A$ let $[x,\,\,\,ay]=x(a)\cdot y+a\cdot
[x,\,\,\,y]$.

Let for any positive integer $n$, $\Omega _K^n(L,\,\,\,A)$ be the space of
skew-symmetric, $K$-multilinear mappings from $L^n$ into $A$, where $K$ is
the field of either real or complex numbers. Using the formula \ref{Formula4}
for the Schouten bracket on $\wedge L=\oplus \wedge ^kL$, we obtain that for
any $u\in \wedge ^mL$, $v\in \wedge ^nL$ and $\omega \in \Omega
_K^{m+n-1}(L,\,\,\,A)$:%
$$
\omega ([u,\,\,\,v])=\omega (\partial (u)\wedge v)+(-1)^m\omega (u\wedge
\partial (v))-\omega (\partial (u\wedge v)) 
$$
or, in the other notations 
\begin{equation}
\label{Formula5}\omega ([u,\,\,\,v])=(-1)^{(m+1)n}i_v\omega (\partial
(u))+(-1)^mi_u\omega (\partial (v))-\omega (\partial (u\wedge v)) 
\end{equation}
where, for $\alpha \in \Omega _K^p(L,\,\,\,A)$ and $x\in \wedge ^qL$, under
the notation $i_x\omega $, we mean the element of the space $\Omega
_K^{p-q}(L,\,\,\,A)$ defined as $(i_x\alpha )(y)=\alpha (x\wedge y)$, for
any $y\in \wedge ^{p-q}L$.

Using the dual notations, the expression \ref{Formula5} can be written in
the following form 
\begin{equation}
\label{Formula6}\omega ([u,\,\,\,v])=(-1)^{(m+1)n}(\partial ^{*}i_v\omega
)(u)+(-1)^m(\partial ^{*}i_u\omega )(v)-(\partial ^{*}\omega )(u\wedge v) 
\end{equation}
where: $(\partial ^{*}\alpha )(x)=\alpha (\partial x)$, for any $\alpha \in
\Omega _K^p(L,\,\,\,A)$ and $x\in \wedge ^pL$.

By the definition of the classical exterior differential, we have that $%
d\alpha =\partial ^{*}\alpha +\partial _1\alpha $, where, for $u_1,\ldots
,u_{p+1}\in L$, the expression $\partial _1\alpha $ is defined as%
$$
(\partial _1\alpha )(u_1\wedge \ldots \wedge
u_{p+1})=\sum_{i=1}^{p+1}(-1)^{i-1}u_i\alpha (u_1\wedge \ldots \wedge 
\widehat{u_i}\wedge \ldots \wedge u_{p+1}) 
$$
It is easy to verify, that%
$$
(-1)^{(m+1)n}(\partial _1i_v\omega )(u)+(-1)^m(\partial _1i_u\omega
)(v)-(\partial _1\omega )(u\wedge v)=0 
$$
Therefore, in the expression \ref{Formula6}, we can replace the operator $%
\partial ^{*}$ by the operator $d$: 
\begin{equation}
\label{Formula7}\omega ([u,\,\,\,v])=(-1)^{(|u|+1)|v|}(di_v\omega
)(u)+(-1)^{|u|}(di_u\omega )(v)-(d\omega )(u\wedge v) 
\end{equation}

If we consider the subspace\thinspace $\Omega (L,\,\,\,A)=\oplus
_{i=0}^\infty \Omega ^i(L,\,\,\,A)$ of the space\\$\Omega
_K(L,\,\,\,A)=\oplus _{i=0}^\infty \Omega _K^i(L,\,\,\,A)$ consisting of the 
$A$-multilinear mappings, the formula \ref{Formula7} for the Schouten
bracket is more convenient then the formulas \ref{Formula4}, \ref{Formula5}
and \ref{Formula6}, as the subspace $\Omega (L,\,\,\,A)$ is invariant under
the action of the operator $d$, and besides that, the formula \ref{Formula7}
can be used as an invariant definition of the Schouten bracket in some
cases. For example, for the covariant, skew-symmetric tensor fields on a
smooth manifold.

As we know, an invariant element $p\in L\wedge L$, defines an operator $%
\partial _p=[p,\,\,\,]:\wedge L\longrightarrow \wedge L$, of degree $+1$,
which is a coboundary operator. The dual operator%
$$
\begin{array}{l}
\partial _p^{*}:\Omega _K(L,\,\,\,A)\longrightarrow \Omega _K(L,\,\,\,A) \\  
\\ 
(\partial _p^{*}\omega )(x)=\omega ([p,\,\,\,x]) 
\end{array}
$$
is an operator of degree $-1$ and is a boundary operator: $\partial
_p^{*}\circ \partial _p^{*}=0$. Using the formula \ref{Formula7} we obtain
the following expression for $\partial _p^{*}$:%
$$
(\partial _p^{*}\omega )(u)=(d\omega )(p\wedge u)-(di_p\omega
)(u)-(-1)^{|u|}(di_u\omega )(p) 
$$
Or, in more brief notations 
\begin{equation}
\label{Formula8}(\partial _p^{*}\omega )(u)=(i_p\circ d-d\circ i_p)(\omega
)(u)-(-1)^{|u|}(di_u\omega )(p) 
\end{equation}
It is clear that the subalgebra $\Omega (L,\,\,\,A)$ of the algebra $\Omega
_K(L,\,\,\,A)$ is not invariant under the action of the operator $\partial
_p^{*}$, as for $\omega \in \Omega ^n(L,\,\,\,A)$, $a\in A$, and $x\in
\wedge ^nL$, we have%
$$
\begin{array}{l}
(\partial _p^{*}\omega )(a\cdot x)=-\omega ([p,\,\,\,a\cdot x])=-\omega ( 
\widetilde{p}(a)\wedge x+a[p,\,\,\,x])= \\  \\ 
=-((-1)^{|x|}(i_x\omega )(\widetilde{p}(a))+a\cdot \omega
([p,\,\,\,x]))=a\cdot (\partial _p^{*}\omega )(x)-(-1)^{|x|}(i_x\omega )(%
\widetilde{p}(a)) 
\end{array}
$$
To ''correct'' the operator $\partial _p^{*}$, so that the algebra of
differential forms $\Omega (L,\,\,A)$ be invariant under its action, we
remove the last term in the \ref{Formula8}. The result is exactly the
boundary operator of the canonical complex for Poisson manifold, which is
well-known in the case when $L$ is the Lie algebra of vector fields on some
Poisson manifold $M$, and $A$ is the commutative algebra of smooth functions
on $M$ (see \cite{brylinski}) 
\begin{equation}
\label{Formula9}\partial :\Omega ^m(L,\,\,\,A)\longrightarrow \Omega
^{m-1}(L,\,\,\,A),\,\,\,\,\partial =i_p\circ d-d\circ i_p 
\end{equation}

For $p$, define the following bilinear mapping: 
\begin{equation}
\label{Formula10}
\begin{array}{l}
p:\Omega ^m(L,\,\,\,A)\times \Omega ^n(L,\,\,\,A)\longrightarrow \Omega
^{m+n-2}(L,\,\,\,A) \\  
\\ 
p(\alpha ,\,\,\,\beta )=i_p(\alpha \wedge \beta )-i_p\alpha \wedge \beta
-\alpha \wedge i_p\beta 
\end{array}
\end{equation}
The Schouten bracket on the anticommutative graded algebra $\Omega
(L,\,\,\,A)=\oplus \Omega ^k(L,\,\,\,A)$ can be defined as 
\begin{equation}
\label{Formula11}[\alpha ,\,\,\,\beta ]=d\,\,p(\alpha ,\,\,\,\beta
)-p(d\,\,\alpha ,\,\,\,\beta )-(-1)^{|\alpha |}p(\alpha ,\,\,\,d\beta ) 
\end{equation}
(see \cite{karasev-maslov}).

\begin{theorem}
The bracket on $\Omega (L,\,\,\,A)$ defined by the formula \ref{Formula11}
coincides with the bracket $[\,\,\,,\,\,\,]_\delta $ which is the deviation
of the operator $\delta $ from antiderivation. That is: for any $\alpha \in
\Omega ^m(L,\,\,\,A)$, and $\beta \in \Omega (L,\,\,\,A)$, the following
equality is true\\$\delta \alpha \wedge \beta +(-1)^m\alpha \wedge \beta
-\delta (\alpha \wedge \beta )=dp(\alpha ,\,\,\,\beta )-p(d\alpha
,\,\,\,\beta )-(-1)^mp(\alpha ,\,\,\,d\beta )$
\end{theorem}

The proof of this theorem consists of simple verifying of the equality
keeping in mind the formulas \ref{Formula9}, \ref{Formula10} and \ref
{Formula11}.

For any $a\in A$, define the element $da\in \Omega ^1(L,\,\,\,A)$, as $%
(da)(X)=X(a)$ for any $X\in L$. Consider the subalgebra of the $\Omega
(L,\,\,\,A)$ generated by $A$ and $dA\subset \Omega ^1(L,\,\,\,A)$. Denote
this subalgebra by $\widetilde{\Omega }(L,\,\,\,A)$, and the corresponding
grading subspaces by $\widetilde{\Omega }^k(L,\,\,\,A)$ for $k=0,\cdots
,\infty $. As it follows from the definition, each $\widetilde{\Omega }%
^k(L,\,\,\,A)$ consists of the elements of the form $\sum%
\limits_{i=1}^na_0^i\,\,d\,a_1^i\wedge d\,a_k^i$. The Poisson bracket on $A$
defined by $p$, as $\{a,\,\,\,b\}=i_p(da\wedge db)$, for $a,\,\,\,b\in A$,
gives the same expression for the operator $\delta $, as in the case when $A$
is the algebra of smooth functions on some Poisson manifold and $L$ is the
Lie algebra of the vector fields on the same manifold (see \cite{brylinski}%
): 
\begin{equation}
\label{Formula12}
\begin{array}{l}
\delta (a_0da_1\wedge \ldots \wedge
a_n)=\sum\limits_{i=1}^n(-1)^{i+1}\{a_0,\,\,\,a_i\}da_1\wedge 
\widehat{da_i}\wedge da_n+ \\ +\dsum\limits_{i<j}(-1)^{i+j}a_0d\{a_i,\,\,%
\,a_j\}\wedge da_1\wedge \ldots \wedge \widehat{da_i}\wedge \ldots \wedge 
\widehat{da_j}\wedge \ldots \wedge da_n 
\end{array}
\end{equation}
By using of this formula, it is easy to verify that on $\widetilde{\Omega }%
(L,\,\,\,A)$ the condition \ref{FormulaX4} for $\delta $ is true, therefore,
the bracket defined by \ref{Formula11} or by $[\alpha ,\,\,\,\beta ]=\delta
\alpha \wedge \beta +(-1)^{|\alpha |}\alpha \wedge \delta \beta -\delta
(\alpha \wedge \beta )$ on $\widetilde{\Omega }(L,\,\,\,A)$ gives a Lie
superalgebra structure, which is the extension of the Lie algebra structure
on $\widetilde{\Omega }^1(L,\,\,\,A)$. So, in the case when $A=C^\infty (M)$
for some Poisson manifold $M$, and $L$ is the Lie algebra of vector fields
on the same manifold, we can state that the supercommutator of differential
forms on $M$ is the deviation of the canonical boundary operator $\delta $
from antidifferential. An element $x\wedge y\in L\wedge L$ defines the
mapping from $A$ into $L$, $a\mapsto U_a$, as $U_a=x(a)\cdot y-y(a)\cdot x$.
It is clear that for each $a\in A$, the expression $U_a$ depends only on $%
da\in \widetilde{\Omega }^1(L,\,\,\,A)$. This mapping can be extended
linearly for any $p\in L\wedge L$. After that, for any $p\in L\wedge L$ we
can define the mappings $\widetilde{p}:\widetilde{\Omega }%
^k(L,\,\,\,A)\longrightarrow \wedge ^kL,\,\,\,k=0,\cdots ,\infty $ as follows%
$$
\widetilde{p}(a_0da_1\wedge \ldots \wedge da_k)=a_0U_{a_1}\wedge \ldots
\wedge U_{a_k} 
$$
For any fixed $\omega \in \widetilde{\Omega }^n(L,\,\,\,A)$, define a
mappings:

$*:\widetilde{\Omega }^k(L,\,\,\,A)\longrightarrow \widetilde{\Omega }%
^{n-k}(L,\,\,\,A)$

as

$*(a_0da_1\wedge \ldots \wedge da_k)=a_0(i_{U_{a_k}}\circ \cdots \circ
i_{U_{a_1}})\omega $.

In the case when $M$ is a symplectic manifold with a symplectic form $\alpha 
$, $A=C^\infty (M)$, $p$ is the bivector field corresponding to the form $%
\alpha $, $L$ is the Lie algebra of vector fields on $M$, and $\omega
=\alpha ^{(\dim M)\,\,/\,\,2}$, the operator $*$ is the well-known analogue
of the star operator on a Riemannian manifold (see \cite{brylinski}).

\begin{theorem}
If $\omega $ satisfies the following conditions

$d\omega =0$

$da\wedge \omega =0$ for each $a\in A$

then the equality $*\delta =(-1)^kd*$ is true on $\widetilde{\Omega }%
^k(L,\,\,\,A)$, if and only if $(d\circ i_{U_a})\omega =0$, for any $a\in A$.
\end{theorem}

{\bf Proof. }on $\widetilde{\Omega }^1(L,\,\,\,A)$ we have:%
$$
\begin{array}{l}
(*\delta )(a_0da_1)=*(\{a_0,\,\,\,a_1\})=\{a_0,\,\,\,a_1\}\cdot \omega ; \\  
\\ 
(d*)(a_0da_1)=d(a_0i_{U_{a_1}}\omega )=da_0\wedge i_{U_{a_1}}\omega
+a_0di_{U_{a_1}}\omega 
\end{array}
$$
Consequently: $(*\delta +d*)(a_0da_1)=\{a_0,\,\,\,a_1\}\cdot \omega
+da_0\wedge i_{U_{a_1}}\omega +a_0di_{U_{a_1}}\omega
=-i_{U_{a_1}}(da_0\wedge \omega )+a_0di_{U_{a_1}}\omega
=a_0di_{U_{a_1}}\omega $.

Therefore, on the space $\widetilde{\Omega }^1(L,\,\,\,A)$ the equality $%
*\delta =-d*$ is true if and only if $(d\circ i_{U_a})\omega =0$ for any $%
a\in A$.

To proof the equality $*\delta =(-1)^kd*$ for each $\widetilde{\Omega }%
^k(L,\,\,\,A)$, the following well-known formula can be used%
$$
\begin{array}{l}
(L_X\omega )(X_1,\cdots ,X_n)=(i_Xd\omega +di_X\omega )(X_1,\cdots ,X_n)= \\
\\ 
=X\omega (X_1,\cdots ,X_n)-\sum_i\omega (X_1,\cdots ,[X,\,\,\,X_i],\cdots
,X_n) 
\end{array}
$$

\bigskip\ 

So, we can state that the operator $*$ induces a homomorphism from the
homology space $H_i(L,\,\,A,\,\,\delta )$ of the complex $(\widetilde{\Omega 
}(L,\,\,\,A),\,\,\,\delta )$ into the cohomology space $H^{n-1}(L,\,\,\,A)$
of the complex $(\widetilde{\Omega }(L,\,\,\,A),\,\,\,d)$.

\section{Brief overview of the geometric structure of Poisson Manifolds}

Further we shall consider the case when the commutative algebra $A$ is the
algebra $C^\infty (M)$ for some smooth manifold $M$, $L$ is the Lie algebra
of vector fields on $M$ and therefore $\Omega (M)$ is the exterior algebra
of differential forms on $M$. An involutive element $p\in V^2(M)$, where $%
V^2(M)$ is the space of the second-order covariant antisymmetric tensor
fields on $M$, defining a Poisson algebra structure on the space $C^\infty
(M)$ is called as a {\em bivector field} on the manifold $M$.

Let $\pi $ be the differential system on $M$ derived by the set of the
vector fields of the type $X_f=\{f,\,\,\,\cdot \}$ for $f\in C^\infty (M)$.
In other words, for any point $x\in M$, the subspace $\pi (x)\subset T_xM$
is defined as%
$$
\begin{array}{l}
\pi (x)=\{\,\,\,u\in T_xM\,\,\,|\,\,\,\beta (u)=(\alpha \wedge \beta )(p_x)
\\ 
\,\,for\,\,some\,\,\alpha \in T_x^{*}M\,\,\,and\,\,\,each\,\,\,\beta \in
T_x^{*}M\,\,\,\} 
\end{array}
$$
The rank of the differential system $\pi $ at any point $x\in M$ (i.e. the
dimension of the space $\pi (x)$) equal to the rank of the bivector field $p$
at the point $x$ ($rank(p_x)=r\,\,\,\Leftrightarrow
\,\,\,r=2k,\,\,\,for\,\,\,some\,\,\,integer\,\,k\,\,\,such\,\,\,that\,\,\,%
\wedge ^kp_x\neq 0\,\,\,and\,\,\,\wedge ^{k+1}p_x=0$).

For any function $f\in C^\infty (M)$, let $\varphi _t,\,\,\,t\in R$ be the
one-parameter group of diffeomorphisms of the manifold $M$, corresponding to
the vector field\\$X_f=\{f,\,\,\,\cdot \}$. The bivector field $p$ is
conserved by the group $\varphi _t$: the latter statement is equivalent to
the following equality%
$$
(d(g\circ \varphi _t)\wedge d(h\circ \varphi _t))(p)\,\,\circ \,\,\varphi
_t^{-1}=(dg\wedge dh)(p) 
$$
which itself, is equivalent to 
\begin{equation}
\label{Formula13}\{g\circ \varphi _t,\,\,\,h\circ \varphi
_t\}=\{g,\,\,\,h\}\circ \varphi _t 
\end{equation}
the latter is a result of the Jacoby identity for the functions $f,\,\,g$,
and $h$, which is the infinitesimal variant of \ref{Formula13}.

It is natural to ask, is the differential system $\pi $ integrable or not.
Note, that it is an involutive system%
$$
\begin{array}{l}
X,\,\,Y\in \pi \Leftrightarrow (X=\sum \varphi _i\{f_i,\,\,\,\cdot
\},\,\,\,Y=\sum \psi _i\{g_i,\,\,\,\cdot \})\Rightarrow  \\  
\\ 
\lbrack X,\,\,\,Y]=\sum (\varphi _i\{f_i,\,\,\,\psi _i\}\cdot
\{g_i,\,\,\,\cdot \}-\psi _i\{g_i,\,\,\,\varphi _i\}\{f_i,\,\,\,\cdot \})+
\\ 
+\sum \varphi _i\psi _i\{\{f_i,\,\,\,g_i\},\,\,\,\cdot \}\Rightarrow
[x,\,\,\,y]\in \pi 
\end{array}
$$
Moreover, the following theorem describes the exact condition for any
bivector field $p$ the corresponding differential system $\pi $ be
involutive:

\begin{theorem}
The differential system $\pi $ is involutive if and only if $%
[p,\,\,\,p]|_x\in \pi |_x\wedge \pi |_x\wedge \pi |_x$ for every point $x\in
M$.
\end{theorem}

{\bf Proof. }To prove the theorem, the following formula is useful:

for $\omega \in \Omega ^2(M),\,\,\,\alpha ,\,\,\beta \in \Omega ^1(M)$ and $%
X,\,\,\,Y\in V^2(M)$%
\begin{equation}
\label{Formula14}
\begin{array}{l}
(\omega \wedge \alpha \wedge \beta )(X\wedge Y)=\omega (X)\cdot (\alpha
\wedge \beta )(Y)+\omega (Y)\cdot (\alpha \wedge \beta )(X)- \\ 
-\omega (\widetilde{X}(\alpha ),\,\,\widetilde{Y}(\beta ))+\omega (%
\widetilde{X}(\beta ),\,\,\widetilde{Y}(\alpha )) 
\end{array}
\end{equation}
It is sufficient to verify this formula in the case when $\omega =\varphi
\wedge \psi $, for any $\varphi ,\,\,\psi \in \Omega ^1(M)$. In this case we
have the following%
$$
\begin{array}{l}
(\varphi \wedge \psi \wedge \alpha \wedge \beta )(X\wedge Y)=(\varphi \wedge
\psi )(X)\cdot (\alpha \wedge \beta )(Y)+ \\ 
+(\varphi \wedge \alpha )(X)\cdot (\beta \wedge \psi )(Y)+(\varphi \wedge
\beta )(X)\cdot (\varphi \wedge \alpha )(Y)+ \\ 
+(\psi \wedge \alpha )(X)\cdot (\varphi \wedge \beta )(Y)+(\psi \wedge \beta
)(X)\cdot (\alpha \wedge \varphi )(Y)+ \\ 
+(\alpha \wedge \beta )(X)\cdot (\varphi \wedge \psi )(Y)= \\ 
=\omega (X)\cdot (\alpha \wedge \beta )(Y)+\omega (Y)\cdot (\alpha \wedge
\beta )(X)- \\ 
-\omega (\widetilde{X}(\alpha ),\,\,\,\widetilde{Y}(\beta ))+\omega (%
\widetilde{X}(\beta ),\,\,\,\widetilde{Y}(\alpha )) 
\end{array}
$$

The statement of the theorem , translated on the language of a local
coordinate system $\{x_1,\ldots ,x_n\}$ is the following: for each $%
i,\,\,j\in \{1,\ldots ,n\}$ the vector field $[\widetilde{p}(dx_i),\,\,\,%
\widetilde{p}(dx_j)]$ takes its values in the differential system $\pi $;
which is the same thing, that $\sigma ([\widetilde{p}(dx_i),\,\,\,\widetilde{%
p}(dx_j)])=0$ for each $\sigma \in (\pi )^{\perp }\subset \Omega ^1(M)$.

Using the formula \ref{Formula7} for the Schouten bracket, we obtain:

$(d\sigma \wedge dx_i\wedge dx_j)(p\wedge p)=(2d\sigma )(p)\cdot (dx_i\wedge
dx_j)(p)-(\sigma \wedge dx_i\wedge dx_j)([p,\,\,p])$.

By using of the formula \ref{Formula14}, we obtain:

$(d\sigma \wedge dx_i\wedge dx_j)(p\wedge p)=(2d\sigma )(p)\cdot (dx_i\wedge
dx_j)(p)-(2\sigma )(\widetilde{p}(dx_i,\,\,\,\widetilde{p}(dx_j)))$.

Hence, we have:

$(\sigma \wedge dx_i\wedge dx_j)([p,\,\,p])=-(2\sigma )(\widetilde{p}%
(dx_i,\,\,\,\widetilde{p}(dx_j)))$.

Recall that $\sigma \in (\pi )^{\perp }$, the latter equality ends the proof
of the theorem.

\bigskip\ 

If the rank of a differential system is constant, then its integrability
follows from the Frobenius's classical theorem; but generally, the
differential system $\pi $, is not of a constant rank. Despite this, the
differential system $\pi $ is always integrable, and it is a result of the
Hermann's theorem (see \cite{hermann}), which is a generalization of the
Frobenius's theorem about the integrability of differential systems of
non-constant rank. The necessary and sufficient condition for the
integrability of a differential system, as it is stated in the above
mentioned theorem, is the conservation of the rank of the system along the
integral paths of this system. This condition is satisfied for the
differential system $\pi $, which follows from the fact that the
one-parameter groups of the Hamiltonian vector field, conserve the bivector
field $p$, and therefore its rank.

An integral leaf of the differential system $\pi $ is called symplectic leaf.

The restriction of the Poisson structure on any integral leaf of the
differential system $\pi $ is non-singular; hence, a symplectic structure is
induced by the bivector field $p$ on such a leaf. Let us denote the
symplectic form induced by the Poisson structure on a symplectic leaf $N$ by 
$\omega _N$. For $x\in N,\,\,\,u\in T_xN$, and $v\in T_xN$ we have that $%
\omega _N(u,\,\,\,v)=\{f,\,\,\,g\}(x)$, where $u=\{f,\,\,\,\cdot \}|_x$, and 
$v=\{g,\,\,\,\}|_x$.

One of the indicators of the singularity of a Poisson structure is the
existence of such smooth function on $M$, which commutes with all functions
on $M$ and is not constant, i.e. the center of the Lie algebra of smooth
functions $Z(M)$, does not coincide to the set of the constant functions.
The elements of the center $Z(M)$ are know as Casimir functions. From the
singularity of the Poisson structure $p$ does not follow the existence of a
non-constant Casimir function.

For instance, if one of the symplectic leaves is everywhere dense in the
manifold $M$ then a Casimir function can be only constant.

By way of illustration, consider the following

{\bf Example}: let $M$ be a two-dimensional symplectic manifold and $p$ be
the corresponding non-singular bivector field on $M$. Let $\varphi $ be a
nonconstant smooth function on $M$. The bivector field $p_1=\varphi \cdot p$%
, is involutive as well as $p$. If the set $\varphi ^{-1}(0)$ is not empty,
the Poisson structure defined by $p_1$, is singular at the points of the set 
$\varphi ^{-1}(0)$, which follows from the relation between the bracket $%
\{\,\,\,,\,\,\,\}_1$ defined by $p_1$ and the bracket $\{\,\,\,,\,\,\,\}$
defined by $p$: $\{f,\,\,\,g\}_1=\varphi \cdot \{f,\,\,\,g\}$, for any $%
f,\,\,\,g\in C^\infty (M)$. If a function $f\in C^\infty (M)$ is a Casimir
function, then we have the following%
$$
\varphi \cdot \{f,\,\,\,\cdot \}=0\Rightarrow \{f,\,\,\,\cdot
\}|_{M\backslash \varphi ^{=1}(0)}=0\Rightarrow f=const 
$$
If $\varphi $ is such, that the set $M\backslash \varphi ^{-1}(0)$ is
everywhere dense (for example, in the case when the set $\varphi ^{-1}(0)\}$
consists only one point $x_0$) then we have that the function $f$ is
constant everywhere on the manifold $M$. So, this is an example of the
situation when a Poisson structure is singular, but Casimir function can be
only constant.

Further, we shall extend (in some sense) the definition of the Poisson
bracket for distributions on a smooth manifold, and be looking for Casimir
functions in the set of distributions.

\section{Distributions on Poisson manifold}

Distribution on a smooth manifold $M$ is a linear function on the subspace
of the space $C^\infty (M)$ consisting of the functions with a compact
support. For simplicity assume that the manifold $M$ is compact, which
implies that a distribution on $M$ is simply a linear function on the space $%
C^\infty (M)$.

Let us denote the space of all distributions on the manifold $M$ by $F(M)$.
Using the classical notations, the value of a distribution $\Phi $ on a
function $\varphi \in C^\infty (M)$, we denote by $<\Phi ,\,\,\varphi >$.

The product of a function $f$, on a distribution $\Phi $, is defined as the
distribution $f\cdot \Phi $ such that, for each $\varphi \in C^\infty
(M):\,\,\,<f\cdot \Phi ,\,\,\varphi >=<\Phi ,\,\,f\cdot \varphi >$. This
operation makes the space $F(M)$ a $C^\infty (M)$-module.

For a vector field $X\in V^1(M)$ and a distribution $\Phi $, the
distribution $X(\Phi )$ is defined as $<X(\Phi ),\;\;\varphi >=-<\Phi
,\,\,X(\varphi )>$. This action makes the $C^\infty (M)$-module, $F(M)$, a $%
V^1(N)$-module. That is: for any $f\in C^\infty (M)$, $X\in V^1(M)$ and $%
\Phi \in F(M)$, we have: $X(f\cdot \Phi )=X(f)\cdot \Phi +f\cdot X(\Phi )$,
which follows from%
$$
\begin{array}{l}
<X(f\cdot \Phi ),\,\,\varphi >=-<f\cdot \Phi ,\,\,X(\varphi )>=-<\Phi
,\,\,f\cdot X(\varphi )>= \\ 
=<\Phi ,\,\,X(f)\cdot \varphi >-<\Phi ,\,\,X(\varphi \cdot f)>= \\ 
=<X(f)\cdot \Phi ,\,\,\varphi >+<X(\Phi ),\,\,\varphi \cdot f>= \\ 
=<X(f)\cdot \Phi ,\,\,\varphi >+<f\cdot X(\Phi ),\,\,\varphi > 
\end{array}
$$

Let $M$ be a Poisson manifold.

As the action of a vector field on a distribution is defined, it can be
defined the Poisson bracket of a function $f$ and a distribution $\Phi
:\,\,\,\{f,\,\,\,\Phi \}=X_f(\Phi )$, where $X_f$ is the Hamiltonian vector
field corresponding to the function $f$. The latter can be written as 
\begin{equation}
\label{Formula15}
\begin{array}{l}
<\{f,\,\,\,\Phi \},\,\,\varphi >=<X_f(\Phi ),\,\,\,\varphi >= \\ 
=-<\Phi ,\,\,\,X_f(\varphi )>=<\Phi ,\,\,\,\{\varphi ,\,\,\,f\}> 
\end{array}
\end{equation}
It makes the $C^\infty (M)$-module $F(M)$ a Lie algebra module over $%
C^\infty (M)$. That is:%
$$
\begin{array}{l}
for\,\,\,f,g\in C^\infty (M)\,\,\,\,and\,\,\,\,\Phi \in F(M), \\  
\\ 
\{f,\,\,\,g\cdot \Phi \}=\{f,\,\,g\}\cdot \Phi +g\cdot \{f,\,\,\Phi \} 
\end{array}
$$
Besides that, we have that, for any fixed $\Phi \in F(M)$, the first order
differential operator $\{\Phi ,\,\,\cdot \}:C^\infty (M)\longrightarrow F(M)$
is such that for any $\varphi ,\psi \in C^\infty (M):\,\{\Phi ,\,\,\varphi
\psi \}=\varphi \{\Phi ,\,\,\psi \}+\psi \{\Phi ,\,\,\varphi \}$. To verify
this, the following expression can be used: $\{\varphi \psi ,\,\,\cdot
\}=\varphi \{\psi ,\,\,\}+\psi \{\varphi ,\,\,\}$.

After we have defined the Poisson bracket of a distribution and a smooth
function on the manifold $M$, note that, if the Poisson structure on $M$ is
singular, but has not a nonconstant center, can have such ''center'' in the
space of the distributions. That is, there can be such a distribution $\Phi
\in F(M)$ that $\{\Phi ,\,\,\varphi \}=0$ for each $\varphi \in C^\infty (M)$%
. In the situation described at the end of the previous section, such
distributions are the Dirac functions $\delta _a$ for any $a\in \varphi
^{-1}(0),\,\,\,\delta _a(f)=f(a)$. In this case, for any $f,g\in C^\infty
(M) $, we have%
$$
\begin{array}{l}
<\{\delta _a,\,\,f\}_1,\,\,g>=<\delta _a,\,\,\{f,\,\,g\}_1>= \\ 
=<\delta _a,\,\,\varphi \{f,\,\,g\}>=\varphi (a)\{f,\,\,g\}(a)=0 
\end{array}
$$

Now, we shall describe some general construction to build a distributions
''commuting'' with each smooth function on the manifold $M$.

Let us recall the following formula for the Poisson bracket of two functions
on a symplectic manifold with a symplectic form $\omega $: 
\begin{equation}
\label{Formula16}\{f,\,\,\,g\}\cdot \omega ^n=n\cdot dg\wedge df\wedge
\omega ^{n-1} 
\end{equation}
where $n$ is the half-dimension of the manifold. The formula is the result
of the following%
$$
\{f,\,\,\,g\}\cdot \omega ^n=i_{U_f}(dg\wedge \omega ^n)+dg\wedge
i_{U_f}\omega =ndg\wedge df\wedge \omega ^{n-1} 
$$

Let $N$ be a symplectic leaf in the Poisson manifold $M$. As it was
mentioned early, the restriction of the bivector field $p$, on the leaf $N$
is not singular and we denote the corresponding symplectic form by $\omega
_N $. Consider the following distribution on the manifold $M$:%
$$
\delta _N:C^\infty (M)\longrightarrow R,\,\,\,\,<\delta _N,\,\,\varphi
>=\dint\limits_N\varphi |_N\cdot \omega ^k 
$$
where $\varphi \in C^\infty (M)$ and $k=\frac 12\dim N$.

\begin{theorem}
For any $\varphi \in C^\infty (M)$, we have $\{\delta _N,\,\,\,\varphi \}=0$.
\end{theorem}

{\bf Proof. }By the definition of the Poisson bracket of a distribution and
a smooth function we have:

for any $\varphi ,\,\,\psi \in C^\infty (M)$%
$$
<\{\delta _N,\,\,\varphi \},\,\,\psi >=<\delta _N,\,\,\{\varphi ,\,\,\psi
\}>=\dint\limits_N\{\varphi ,\,\,\psi \}|_N\cdot \omega _N^k 
$$
Keeping in mind the fact that, the Hamiltonian vector fields are tangent to
the symplectic leaves, the formula \ref{Formula16} and the Stokes formula,
we obtain%
$$
\begin{array}{l}
\dint\limits_N\{\varphi ,\,\,\psi \}|_N\cdot \omega
_N^k=\dint\limits_N\{\varphi |_N,\,\,\psi |_N\}\cdot \omega _N^k= \\ 
=n\dint\limits_Nd\,\psi \wedge d\varphi \wedge \omega
_N^{k-1}=n\dint\limits_{\partial N}\,\psi \wedge d\varphi \wedge \omega
_N^{k-1}=0 
\end{array}
$$

\bigskip\ 

Let $F_0(M)$ be the subspace of the space $F(M)$ consisting of the
distributions commuting with every smooth function; $H_0(M,\,\,\delta )$ be
the space of $0$-dimensional homologies of the canonical complex of the
Poisson manifold $M$, denoted by $(\Omega (M),\,\,\,\delta )$; and $%
H_0(M,\,\,\delta )^{*}$ be the space of linear functions on the space $%
H_0(M,\,\,\delta )$.

\begin{lemma}
The spaces $F_0(M)$ and $H_0(M,\,\,\delta )^{*}$ are isomorphic.
\end{lemma}

{\bf Proof. }As it follows from the definition of the Poisson bracket of a
distribution and a smooth function, the space $F_0(M)$ can be defined as

$$
\begin{array}{l}
F_0(M)=\{\Phi \in F(M)\,\,|\,\,<\Phi ,\,\,\,\{f,\,\,\,g\}>=0 \\  
\\ 
for\,\,\,every\,\,\,f,\,g\in C^\infty (M)\} 
\end{array}
$$

In other words, $F_0(M)=\{C^\infty (M),\,\,\,\,C^\infty (M)\}^{\bot }$, where%
\\$\{C^\infty (M),\,\,\,\,C^\infty (M)\}$ is the space of the sums of the
type%
$$
\dsum \{\varphi _i,\,\,\,\psi _i\},\,\,\,\varphi _i,\,\,\,\psi _i\in
C^\infty (M). 
$$

As it follows from the formula \ref{Formula12} for the canonical coboundary
operator $\delta :\Omega (M)\longrightarrow \Omega (M)$, its action on the
form $\alpha =\dsum \varphi _id\psi _i\in \Omega ^1(M)$ is $\delta (\alpha
)=\dsum \{\varphi _i,\,\,\,\psi _i\}$. Therefore, $\delta (\Omega
^1(M))=\{C^\infty (M),\,\,\,\,C^\infty (M)\}$. As $H_0(M,\,\,\,\delta
)=C^\infty (M)/\delta (\Omega ^1(M))$, we obtain that $\delta (\Omega
^1(M))^{\bot }=H_0(M,\,\,\,\delta )^{*}$

\bigskip\ 

\begin{corollary}
If $M$ is a symplectic manifold, then the space $F_0(M)$ is one-dimensional
and the functional $\delta _\omega $ defined as $<\delta _\omega
,\,\,\,\varphi >=\dint\limits_M\varphi \cdot \omega ^n$, where $\varphi \in
C^\infty (M),\,\,\,\omega $ is the symplectic form and $\dim M=2n$, forms
its basis.
\end{corollary}

{\bf Proof. }If $M$ is a symplectic manifold, then the mapping\\$%
*:H_0(M,\,\,\,\delta )\longrightarrow H^{2n}(M)$, where $H^{2n}(M)$ is the $%
2n$-dimensional De-Rham cohomology space of $M$, is isomorphism. As $M$ is
symplectic, it is an oriented manifold, therefore $H^{2n}(M)\cong R$

\bigskip\ 

Let $N$ be a symplectic leaf in the Poisson manifold $M$, and\\$r:C^\infty
(M)\longrightarrow C^\infty (M)$ be the restriction mapping. It is clear
that $\delta _N=r^{*}(\delta _{\omega _N})$, where $r^{*}:F(M)%
\longrightarrow F(M)$ is the dual mapping, and $\omega _N$ is the symplectic
form on $N$ induced by the Poisson structure. If the mapping $r$ is an
epimorphism, then $Image(r^{*})=(I_N\,)^{\bot }$, where $I_N\,$ is the ideal
of the functions on $M$ vanishing on the submanifold $N$, and $(I_N\,)^{\bot
}$ is its orthogonal subspace in the space $F(M)$.

\begin{theorem}
If a symplectic leaf $N$ in the Poisson manifold $M$ is such that the
restriction mapping $r:C^\infty (M)\longrightarrow C^\infty (M)$ is
epimorphic, then the space $(I_N\,)^{\bot }\cap F_0(M)$ is one-dimensional
and the element $\delta _N$ forms its basis.
\end{theorem}

{\bf Proof. }As the mapping $r$ is a Poisson mapping, i.e. for each pair $%
\varphi ,\,\,\psi \in C^\infty (M):\,\,\pi (\{\varphi ,\,\,\psi \})=\{\pi
(\varphi ),\,\,\pi (\psi )\},\,\,\pi ^{-1}((I_N\,)^{\bot }\cap
F_0(M))=F_0(N) $, which is one-dimensional according to the {\bf Corollary 1}
(see {\bf Lemma 2})

\bigskip\ 

\bigskip\

\end{document}